\documentclass[aps,pre,twocolumn,showpacs]{revtex4}
\usepackage{epsfig}
\usepackage{times}
\begin{document}

\title{Competition of individual and institutional punishments in spatial public goods games}
\author{Attila Szolnoki,$^1$ Gy{\"o}rgy Szab{\'o},$^1$ and Lilla Czak{\'o}$^2$}
\affiliation{$^1$ Research Institute for Technical Physics and Materials Science, P.O. Box 49, H-1525 Budapest, Hungary \\
$^2$ Roland E{\"o}tv{\"o}s University, Institute of Physics, P{\'a}zm{\'a}ny P. s{\'e}t{\'a}ny 1/A, H-1117 Budapest, Hungary}

\begin{abstract}

We have studied the evolution of strategies in spatial public goods games where both individual (peer) and institutional (pool) punishments are present beside unconditional defector and cooperator strategies. The evolution of strategy distribution is governed by imitation based on random sequential comparison of neighbors' payoff for a fixed level of noise. Using numerical simulations we have evaluated the strategy frequencies and phase diagrams when varying the synergy factor, punishment cost, and fine. Our attention is focused on two extreme cases describing all the relevant behaviors in such a complex system. According to our numerical data peer punishers prevail and control the system behavior in a large segments of parameters while pool punishers can only survive in the limit of weak peer punishment when a rich variety of solutions is observed. Paradoxically, the two types of punishment may extinguish each other's impact resulting in the triumph of defectors. The technical difficulties and suggested methods are briefly discussed.  

\end{abstract}

\maketitle

\section{Introduction}
\label{intro}

The emergence of cooperation among selfish individuals is an important and intensively studied puzzle inspired by systems of biology, sociology, or economics \cite{nowak_06, szabo_pr07}. One of the frequently used framework to catch the conflict of individual and common interests is the so-called public goods game (PGG) in which several players decide simultaneously whether they contribute or not to the common venture. The collected income is multiplied by a factor (representing the advantage of collective actions) and shared equally among all members of the group independently of their personal act. Accordingly, defectors, who deny to contribute but enjoy the common benefit (due to the cooperators), collect higher individual payoffs and are favored leading to the ``tragedy of commons'' state \cite{hardin_g_s68}.

In the last decade several mechanisms have already been identified that help resolve this dilemma by ensuring competitive payoff for altruistic (cooperative) players \cite{hauert_s02, nowak_s06, hauert_s07, hauert_bt08, wakano_pnas09, cao_xb_pa10, zhang_jl_pa10, rong_pre10, cheng_hy_njp10, perc_pone10, lin_yt_pa11, cheng_hy_njp11, peng_d_epjb10, du_f_dga11, suzuki_r_ijbic11, ohdaira_acs11}. A plausible idea is to punish defectors by lowering their income which decreases their popularity \cite{fehr_aer00, fehr_n02, sigmund_tee07, sigmund_10}. To punish cheaters, however, can be executed in two significantly different ways. 

Firstly, players can retaliate individually by paying extra cost of punishment as often as they face with defectors. Naturally, this so-called peer punisher strategy fares equally well with pure cooperators in the absence of cheaters. The pure cooperators, however, who do not contribute to the sanctions but utilize the advantage of punishment, can be considered as ``second-order free-riders'' \cite{panchanathan_n04}. As a conclusion, the generally less favored peer punisher strategy will become extinct gradually and the original problem emerges again. Without introducing further complexity this problem cannot be solved in well-mixed population. In structured population, however, an adequate solution may be achieved by utilizing spatial effects \cite{nowak_s06, szabo_pr07}. Here the pure cooperators and peer punishers are able to separate from each other and fight independently against defectors. Since punishers do it more successfully, they eventually displace the pure cooperators via an indirect territorial fight \cite{helbing_ploscb10, helbing_njp10}.

The alternative way to impose sanctions is when players invest a permanent cost into a punishment pool and punish defectors ``institutionally''. In this case, if there is punishment in the group, the fine imposed on defectors may not necessarily depend on the actual number of punishers in the group and the cost of punishers can also be independent on the number of cheaters among group members. In this way the cost is always charged independently of the necessity or efficiency of punishment. In well-mixed population pool punishment can only prevail if ``second-order punishment'' is allowed, i.e. pure cooperators, who do not invest extra cost into the punishment pool, are also fined \cite{sigmund_n10, sigmund_dga11}. In the absence of the latter possibility defectors will spread if the participation in PGG is compulsory. In agreement with the expectations the spatial models offer another type of solutions where the pool punisher strategy can survive without assuming additional punishment of pure cooperators. In the latter case a self-organizing spatio-temporal pattern can be observed \cite{szolnoki_pre11}. The emergence of spatial patterns, maintained by cyclic dominance among three strategies, is a general phenomenon and occurs for a wide variety of systems including PGG \cite{hauert_s02, szabo_prl02, szolnoki_epl10} and different variants of prisoner's dilemma game \cite{szabo_pre02b, hutson_prsb95, reichenbach_jtb08, frey_pa10, szolnoki_pre10b}. 

We note that many aspects of punishment were already investigated in human experiments \cite{clutton_brock_n95, fehr_n02, fehr_n03, semmann_n03, de-quervain_s04, henrich_s06, sasaki_prsb07, egas_prsb08}, as well as by means of mathematical models with three \cite{hauert_s02, bowles_tpb04, brandt_pnas05}, four \cite{sigmund_pnas01, ohtsuki_n09}, and even more strategies \cite{henrich_jtb01, dreber_n08}.

The seminal work of Sigmund {\it et al.} has revealed that pool punishers always lose and peer punishers prevail for well-mixed populations in the absence of second-order punishment \cite{sigmund_n10,sigmund_dga11}. In the present paper we study the competition of punishing strategies by assuming structured population. It will be demonstrated that the stable solution depends sensitively on the relative cost that punishing strategies bear. Accordingly, we have studied two extreme cases illustrating the possible relations of pool and peer punisher players. It should be stressed that in our model additional strategies, such as voluntary optional participation in PGG or second-order punishment of pure cooperators, are not allowed. Despite its simplicity the spatial model exhibits really complex behavior including different space and time scales in connection to the emerging solutions. 

The remainder of this paper is organized as follows. In the next section we describe the studied models by supplying motivations to the suggested extreme cases termed as ``hard'' and ``weak'' peer punishment limits. In Sec.~\ref{costly} we present the solutions obtained by Monte Carlo (MC) simulations for an expensive peer strategy. The results of the other extreme case are presented in Sec.~\ref{cheap}. Finally, we summarize our observations and discuss their potential implications.

\section{Spatial public goods game with punishing strategies}
\label{models}

To preserve comparability with previous works \cite{helbing_njp10,szolnoki_pre11} the public goods game is staged on a square lattice using periodic boundary conditions. We should emphasize, however, that the observed results are robust and are valid in a wider class of two-dimensional lattices. The players are arranged into overlapping five-person ($G=5$) groups in a way that each player at site $x$ serves as a focal player in the group formed together with his/her four nearest neighbors \cite{szolnoki_pre09c, shi_dm_pa09, liu_rr_pa10, shi_dm_epl10, chen_xj_epl10}. Consequently, each individual belongs to $G=5$ different groups and plays five five-person games by following the same strategy in every group he/she is affiliated with. 

According to the four possible strategies, a player on site $x$ is designated as a defector ($s_x = D$), or pure cooperator ($s_x = C$), or peer ($s_x = E$), or pool punisher ($s_x = O$). For the last three strategies the player contributes a fixed amount (equals to $1$ without loss of generality) to the public goods while defectors contribute nothing. The sum of all contributions in each group is multiplied by the factor $r$ ($1<r<G$), reflecting the synergetic effects of cooperation, and the multiplied investment is divided equally among the group members irrespective of their strategies.

In addition to the basic game, defectors may be punished if there are pool or peer punisher players in the group. Pool punishment requires precursory allocation of resources, that is, each punisher contributes a fixed amount $\gamma$ to the punishment pool irrespective of the strategies in its neighborhood. Furthermore, because of the institutional character of this sanction, the resulting $\beta$ fine of defectors is independent on the frequency of pool punishers: the only criterion is to presence at least one pool punisher in the group. 

The character of peer and pool punishments differ significantly. Namely, the cost of peer punishment is charged only if a peer punisher faces with a defector but this cost is multiplied by the number of defectors ($N_{D}^g$) in the given group $g$ ($g=1, \ldots, G$). The latter fact reflects that a peer punisher should penalize every defectors individually. In addition, the fine of a defector originated from peer punishment is accumulated and is proportional to the number of peer punishers $N_{E}^g$ in the group. Denoting the number of cooperators and pool punishers by $N_{C}^g$ and $N_{O}^g$ in the group, the payoff for the possible strategies can be given as
\begin{eqnarray}
\label{eq:payoff}
P_{C}^g&=&r(N_{C}^g+N_{O}^g+N_{E}^g+1)/G - 1, \nonumber \\
P_{O}^g&=&P_{C}^g - \gamma , \\ 
P_{E}^g&=&P_{C}^g - \gamma m N_{D}^g, \nonumber \\
P_{D}^g&=&r(N_{C}^g+N_{O}^g+N_{E}^g)/G -\beta m N_{E}^g -\beta {f(N_{O}^g)} \,, \nonumber
\end{eqnarray}
where the step-like function $f(Z)$ is $1$ if $Z>0$ and $0$ otherwise. The total payoff of a player at site $x$ is accumulated from five public goods games, consequently, $P_{s_x} = \sum_g P_{s_x}^g$ ($g=1, \ldots, G$).

The parameter $m$ in Eqs.~(\ref{eq:payoff}) allows us to quantify two relevant limits in the relation of pool and peer punisher strategies. At the ``hard'' limit of peer punishment ($m=1$) the pool punisher pays a lump cost $\gamma$ while the peer punisher is charged by the same cost $\gamma$ for each action of punishment, that is their corresponding income is reduced by $\gamma N_{D}^g$. Notice, that in spite of their high cost the peer punisher may overcome pool punishers in the absence of defectors. The latter constellations may become relevant in the spatial systems if defectors are present rarely. On the other hand, the hard peer punishment reduces the income of defectors more efficiently if several neighbors apply this strategy against a defector.

The "weak" limit of the peer punishment will be studied at the parameter value $m = 1/(G-1)$. In this case the cost of a pool punisher always exceeds the cost of a peer punisher excepting the case when every group member chooses defection around the $E$ player. Now we consider only the case when the efficiency ({\it i.e.}, corresponding fine) of peer punishment is also reduced by the factor $m$. The above situations raise many questions about the competition and coexistence of the basically different types of punishment.

Following the traditional concept of evolutionary game theory, the population of the more successful individual strategies expands at the disadvantage of others having lower income (fitness). For networked population this strategy update is usually performed via a stochastic imitation of the more successful neighbors. Accordingly, during an elementary step of Monte Carlo simulation a randomly selected player $x$ plays public goods games with her all co-players in $G$ groups and collects  $P_{s_x}$ total payoff as described in Eqs.~(\ref{eq:payoff}). Next, player $x$ chooses one of its four nearest neighbors at random, and the chosen co-player $y$ also acquires its payoff $P_{s_y}$ in the same way. Finally, player $x$ imitates the strategy of player $y$ with a probability $w(s_x \to s_y)=1/\{1+\exp[(P_{s_x}-P_{s_y})/K]\}$, where $K$ quantifies the uncertainty in strategy adoptions \cite{szabo_pre98, szabo_pr07}. 
Generally, the possibility of error in strategy update prevents the system from being trapped in a frozen, metastable state. 
For the sake of direct comparison with previous results \cite{helbing_njp10, szolnoki_pre11} we set $K=0.5$. It is emphasized that the found solutions are robust and remain valid at other (low) values of noise parameter. 

The frequencies of pool and peer punishers ($\rho_{O}$ and $\rho_{E}$), cooperators ($\rho_{C}$) and defectors ($\rho_{D}$) [satisfying the condition $\rho_D+\rho_C+\rho_O+\rho_E=1$] are determined by averaging over a sampling time $t_s$ after a sufficiently long relaxation time $t_r$. The time is measured in the unit of Monte Carlo step (MCS) giving a chance once on average for the players to adopt one of the neighboring strategies. Depending on the values of the parameters $\gamma, \beta$, and $r$ the emerging spatial patterns exhibit a large variety in the characteristic length and time scales. In order to achieve an adequate accuracy (typically the line thickness) we need to vary the linear system size from $L=400$ to $7200$ for sufficiently long sampling and relaxation times (in some crucial cases $t_r=t_s>10^6$ MCS). As we will describe in detail in the subsequent sections the usual choice of random distribution of strategies as an initial state was not always appropriate to find the solution that is valid in the large system size limit. At some parameter values even the largest attainable system size, ($L=7200$),
was not large enough to reach the most stable solution from a random initial state. This problem is related to the fact that the formation of some solutions is characterized by different time scales and the fast relaxation from a random state toward an intermediate (unstable) state prevents the more complex solutions to emerge. In the latter cases we had to use prepared (artificial) initial state ({\it e.g.}, a patch-work-like pattern) combining solutions of subsystems where several strategies are missing. 

\section{Hard peer punishment}
\label{costly}

First we discuss the case of hard peer punishment because it yields simpler phase diagrams. In this case the cost of peer punishers exceeds the cost of pool punishers when several defectors are present in their neighborhood. At the same time the peer punishers can help each other if they form compact colonies in the spatial system and these collaborations multiply the fine reducing the income of neighboring defectors. To reveal the possible stable solutions we have studied different values of synergy factor $r$ exhibiting significantly different results in simpler models studied previously \cite{helbing_njp10, szolnoki_pre11}. The applied values ($r=3.8, 3.5$, and $2$) represent three different classes in the stationary behavior.

The highest synergy factor ($r=3.8$) allows pure cooperators to survive even in the absence of punishment. At a slightly lower synergy value ($r=3.5$) defectors would prevail without punishment, however, both types of punishment (as a possible third strategy) can boost cooperation as it was already shown \cite{helbing_njp10, szolnoki_pre11}. In case of the lowest synergy factor ($r=2$), the simpler three-strategy models predict significantly different behaviors when applying only peer or pool punishment. For low cost values cooperators were unable to survive for the case of peer punishment in the presence of a weak noise allowing additional rare creation of defectors. On the contrary, for pool punishment, the $D$, $C$, and $O$ strategies formed a self-organizing spatial pattern maintained by cyclic dominance. Now the numerical analysis is extended for higher values of $\beta$ and $\gamma$. As a result, we have observed the coexistence of $D$, $C$, and $E$ strategies via a curious mechanism within a region of parameters (not yet investigated previously).

\begin{figure}[ht]
\centerline{\epsfig{file=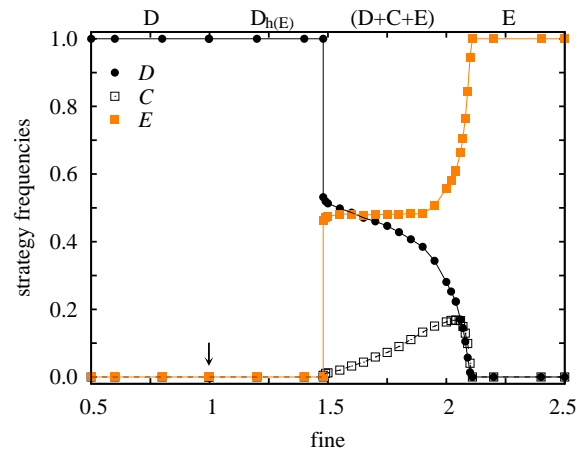,width=7.6cm}}
\caption{(Color online) Average strategy frequencies {\it vs.} fine in the final stationary state for 
hard peer punishment limit ($m=1$) at
$\gamma=0.8$ and $r=2$. The corresponding phases are denoted at the top. Lines are just to guide the eye. The arrow points to the value of fine separating the phases D and D$_h$ where the average invasion velocity between the domains of $D$ and $E$ strategies becomes zero.}
\label{rhos}
\end{figure}

MC simulations were performed to determine the stationary frequency of strategies when varying the value of fine $\beta$ for different values of cost $\gamma$ and $r$. The numerical data indicated 
discontinuous (first-order) or continuous (second-order)
phase transition(s) between phases characterized by basically different compositions and/or spatio-temporal structures as illustrated in Fig.~\ref{rhos} for the lowest value of $r$ we first study. 

\begin{figure}[ht]
\centerline{\epsfig{file=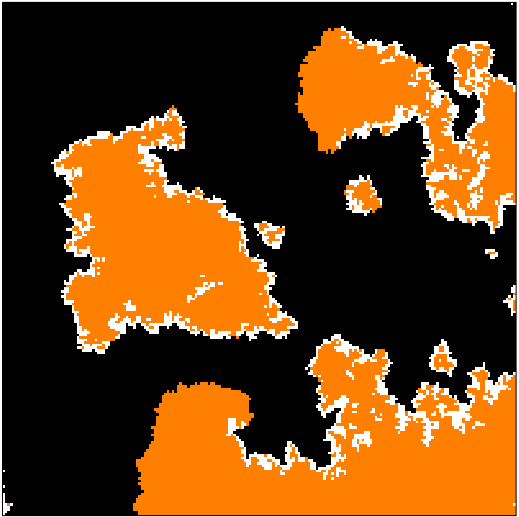,width=7cm}}
\caption{(Color online) Typical arrangement of cooperators (white), defectors (black) and peer punishers (orange - light gray) for the (D+C+E) phase within a $200 \times 200$ part of a larger system at $r=2$, $\gamma=1$, and $\beta=2.5$ in the hard peer punisment limit.
}
\label{snsh1}
\end{figure}

If the system is started from a random initial state with four strategies then the system evolves into the homogeneous (absorbing) state D where only defectors remain alive ($\rho_D=1$) if the fine is smaller than a threshold value $\beta_{th}(\gamma=0.8,r=2,K=0.5)=0.997(4)$ [indicated by an arrow in Fig.~\ref{rhos}]. The simulations show clearly that defectors invade the territories of peer punishers if $\beta < \beta_{th}$. 
For $\beta_{th}< \beta < \beta_{c1}$ the 
superiority
of defectors ($\rho_D=1$) is due to a mechanism that can be understood by considering first the curious coexistence of the $D$, $C$, and $E$ strategies occuring for $\beta_{c1} < \beta < \beta_{c2})$ [$\beta_{c1}(\gamma=0.8,r=2,K=0.5)=1.48(1)$ and $\beta_{c2}(\gamma=0.8,r=2,K=0.5)=2.10(1)$]. The corresponding phase is denoted as (D+C+E). Within this phase strategy $E$ can invade the territories of $D$s along the interfaces separating them as illustrated in the snapshot in Fig.~\ref{snsh1}.
For sufficiently high values of $\beta$ and $\gamma$, however, the expensive action of punishment reduces the income of both defectors and peer punishers along the interface where players can increase their payoff by choosing cooperation. As a result cooperators can spread along these interfaces by forming a ``monolayer''. At the same time the interfacial cooperators serve as a ``cooperator reservoir'' from where cooperation can spread into the phase E via the mechanism described by the voter model \cite{clifford_bm73, cox_ap83, liggett_85}. Rarely the cooperators aggregate in the vicinity of the interface and the given territory becomes unprotected against the invasion of defectors. Consequently, the presence of cooperators along the D-E interfaces reverses the direction of invasion. In the snapshot of Fig.~\ref{snsh1} one can observe both types of invasions balanced in the (D+C+E) phases. 

\begin{figure}[ht]
\centerline{\epsfig{file=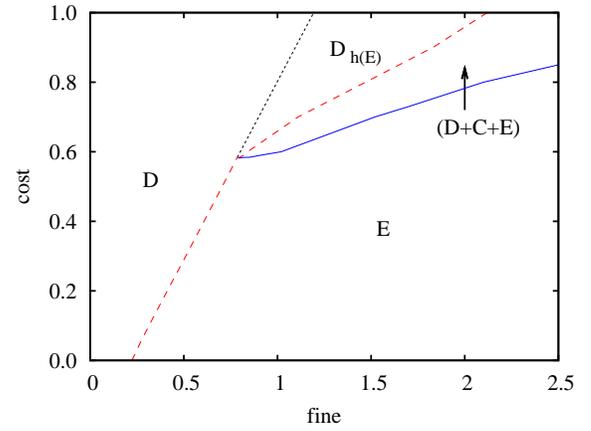,width=7.6cm}}
\caption{(Color online) Cost-fine phase diagram in the hard peer punishment limit ($m=1$) for a low synergy factor ($r=2$). The dashed (red)  and solid (blue) lines represent first- and second-order phase transitions, dotted (black) line separates the homogeneous phases of defectors (D) with different stabilities.}
\label{B1}
\end{figure}

The spreading of cooperators along the D-E interfaces is influenced by the values of $\gamma$ and $\beta$ and it may become so efficient that $C$ monolayers are formed throughout these interfaces. In that case the E domains are invaded by defectors with the assistance of cooperators. Having the last peer punishers removed the defectors sweep out cooperators, too. This process is resembling a real life situation referred as "The Moor has done his duty, the Moor may go". Such a scenario occurs within the phase D$_h$ where subscript h refers to homoclinic instability. The mentioned transient process to D is confirmed by MC simulations for most of the runs in small systems (e.g., $L < 400$). The present system, however, can evolve into the homogeneous state E ($\rho_E=1$) with a probability increasing with $L$. The phase E can conquer D (via a nucleation mechanism) if a small colony of $E$ players survive the extinction of cooperators and the colony size exceeds a critical value during the stochastic evolutionary steps. It is emphasized that the $E$ invasion can be reversed by the offspring of a single cooperator substituted for one of the players along the D-E interface and finally the system evolves into a state prevailed by defectors. 
Notice that pool punishers die out for all the cases plotted in Fig.~\ref{rhos}. Furthermore, the (D+C+E) phases transform into E with a continuous extinction of both the $D$ and $C$ strategies when approaching $\beta_{c2}$. 
Similar numerical investigations are made for many other values of cost $\gamma$ and the results are summarized in a phase diagram plotted in Fig.~\ref{B1}).  

The simulations indicate that both defectors and pool punishers die out within a transient time for sufficiently high values of $\beta$ if $\gamma < \gamma_c(r=2)=0.59(1)$. As the surviving cooperators and peer punishers receive the same payoff therefore the resultant two-strategy evolutionary process becomes equivalent to those described by the voter model. The two-dimensional voter model exhibits an extremely (logarithmically) slow evolution toward one of the (homogeneous) absorbing states \cite{dornic_prl01}. The coexistence of $C$ and $E$ strategies, however, can be destroyed by introducing defectors (as mutants even for arbitrarily small rates) favoring and accelerating the fixation in the homogeneous state of $E$ strategy \cite{helbing_pre10c}. This is the reason why the final stationary state is denoted by E in the phase diagrams throughout the whole paper (see e.g., Figs.~\ref{rhos} and \ref{B1}). Finally we mention that the dotted line in Fig.~\ref{B1} is the analytical continuation of the dashed (red) one separating the phases D and E. Along these lines the average velocity of invasion between the phases E and D becomes zero.

\begin{figure}[ht]
\centerline{\epsfig{file=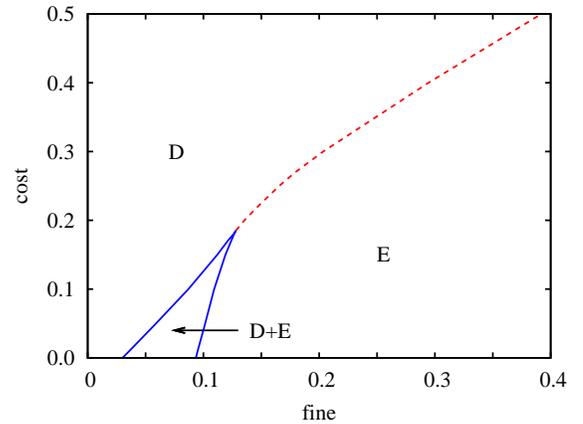,width=7.6cm}}
\caption{(Color online) Cost-fine phase diagram at $m=1$ for $r=3.5$. Solid (blue) and dashed (red) lines represent second-order and first-order phase transitions, respectively. D+E denotes a phase with coexisting $D$ and $E$ strategies.}
\label{B2}
\end{figure}

As expected, the increase of $r$ supports the maintenance of cooperation. Consequently, a smaller fine is capable to suppress defection. As well as for $r=2$ pool punishers die out quickly if $r=3.5$. For high values of $\beta$ and $\gamma$ the cooperators prefer staying along the interfaces separating domains of D and E phases (as described above) and yield a slower tendency toward the final stationary state. The undesired technical difficulty is reduced significantly for lower values of cost and fine where Fig.~\ref{B2} illustrates a discontinuous (first-order) phase transition between the phases D and E at a threshold value of fine increasing with the cost of peer punishment if these quantities exceed the suitable critical values ($\beta_c=0.13(1)$ and $\gamma_c=0.19(1)$ for $r=3.5$ and $K=0.5$).  
When increasing $\beta$ for $\gamma< \gamma_c$ the first order phase transition from the homogeneous D state to E is separated by a coexistence region of $D$ and $E$ strategies. Within this phase the frequency of peer punishers varies continuously from 0 to 1 and both transitions exhibit the general features of directed percolation universality class in agreement with previous results obtained for imitation dynamics \cite{szabo_pre98,chiappin_pre99,szabo_pre02b}.  

\begin{figure}[ht]
\centerline{\epsfig{file=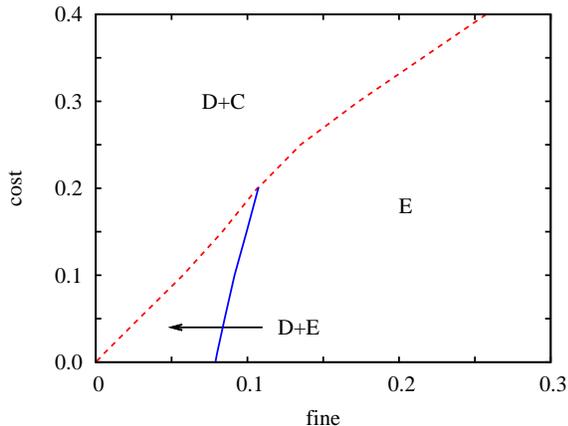,width=7.6cm}}
\caption{(Color online) Cost-fine phase diagram at $m=1$ for $r=3.8$. Phases and phase boundaries are denoted as in Fig.~\ref{B2}.}
\label{B3}
\end{figure}

For higher synergy factors (e.g., $r=3.8$) the cooperators survive in the absence of punishment ($\gamma=0$). Consequently, the homogeneous D phase is missing in the phase diagram. In Figure~\ref{B3} the phase D+C refers to the coexistence of cooperators and defectors in the final stationary states. 
The increase of fine yields a discontinuous transition from D+C to D+E if $\gamma < 0.21(1)$ for the given parameters, otherwise one can observe a first-order transition from D+C to E (within the region of $\gamma$ and $\beta$ plotted in Fig.~\ref{B3}). Within the D+E phase the density of defectors vanishes continuously when approaching the phase boundary separating the phases D+E and E.

\section{Weak peer punishment}
\label{cheap}

In this section we focus on the opposite limit where during the sanction of punishment peer punishers have less cost and enforce lower fine in comparison with those of pool punishers. Using $m = 1/(G-1)$ parameter value, their costs are equal only if a peer punisher is surrounded only by defectors ($N_{D}^g=G-1$). According to a naive argument, the peer punishers might benefit from the powerful fine of pool punishers which strengthen their position further comparing to the latter strategy. This is expected especially after the experience what we observed in the previous section where peer punisher players prevail the system despite of their large extra cost. Following the established protocol, we explore the possible solutions at three representative synergy factors.  

\begin{figure}[ht]
\centerline{\epsfig{file=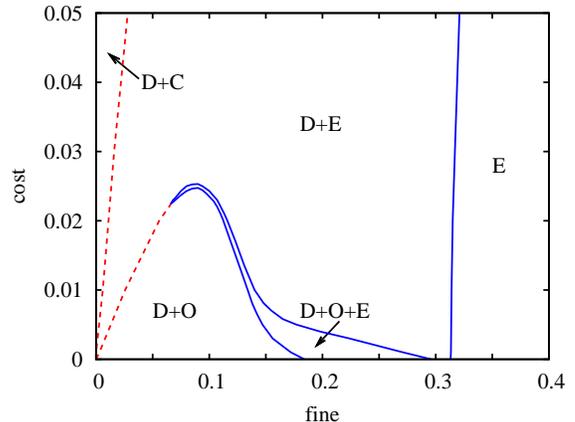,width=7.6cm}}
\caption{(Color online) Phase diagram for the weak peer punishment limit ($m = 1/(G-1)$) at $r=3.8$. Solid (blue) and dashed (red) lines represent second- and first-order phase transitions.}
\label{A_R3_8}
\end{figure}

At high ($r=3.8$) synergy factor the phase diagram, plotted in Fig.~\ref{A_R3_8}, partly supports our expectation. Namely, at high cost ($\gamma>0.0253$) the solutions become identical to those obtained in the absence of $O$ strategies. At low values of cost, however, the above mentioned belief is broken because pool punishers can survive despite that they are charged by a larger permanent cost of punishment. Notice furthermore, that they can fully displace not only pure cooperators but also peer punishers who both can be considered as second-order free-riders. 

\begin{figure}[ht]
\centerline{\epsfig{file=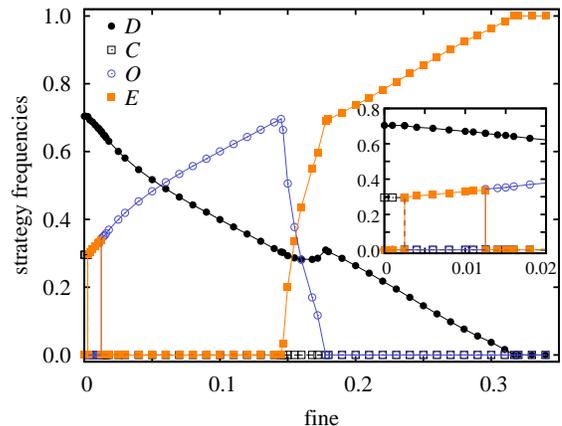,width=7.6cm}}
\caption{(Color online) Strategy frequencies as a function of fine 
in the week peer punishment limit ($m = 1/(G-1)$)
for a low punishment cost ($\gamma=0.005$) 
and $r=3.8$. Inset features the enlargement of the small-fine area.}
\label{indirect}
\end{figure}

Figure~\ref{indirect} shows the variation of strategy frequencies and illustrates five consecutive phase transitions (at $\beta_{c1}$, $\beta_{c2}$, ..., $\beta_{c5}$) when the fine is increased at $\gamma=0.005$. 

The visualization of the time-dependence of spatial strategy distribution has helped us understand what happens and the characteristic mechanisms can be summarized as follows. If the system is started from a random initial state then after a short relaxation process we can observe a sea of defectors with homogeneous islands of cooperative strategies ($C$, $O$, and $E$) for low values of $\beta$. Due to the stochastic dynamics the islands grow and shrink at random and sometimes they can disappear, unite, or split into two. In the late stage of the evolutionary process the pattern formation can be considered as a competition among three two-strategy associations (denoted as D+C, D+O, and D+E) representing the corresponding stationary solutions of subsystems where only two strategies take place \cite{szabo_pr07}. Evidently, the D+C solution can invade the other two associations for infinitesimally small values of fine, because $C$s are not charged by the cost of punishment. The increase of fine, however, favors the survival of the $O$ and $E$ strategies. As a result, the average frequency of the punishing strategies  ($\rho_O$ and $\rho_E$) increases with the fine in the corresponding two-strategy phases (D+O and D+E) while $\rho_C$ remains constant in the phase D+C. The mentioned variations modify the relationship among the three two-strategy solutions. The MC simulations indicate that the D+E phase conquers the whole system if $\beta_{c1} < \beta < \beta_{c2}$ and the D+O phase can be observed in the final state if $\beta_{c2} < \beta < \beta_{c3}$. In the latter two phases the frequency of defectors decreases monotonously with the fine and the punisher islands are simultaneously separated by channels becoming narrower. In parallel with this process the punishing islands unite more frequently enforcing the relevance of direct competition between $E$ and $O$ that boosts the spreading of $E$. The latter effect helps peer punishers to survive in the three-strategy phase D+O+E within the region $\beta_{c3} < \beta < \beta_{c4}$. In the following region of fine ($\beta_{c4} < \beta < \beta_{c5}$) the direct $E$ invasion sweeps out all the pool punishers and the system develops into the phase D+E where the defector frequency approaches 0 at $\beta_{c5}$. If $\beta > \beta_{c5}$ then the system evolves into the phase E as detailed above.

\begin{figure}[ht]
\centerline{\epsfig{file=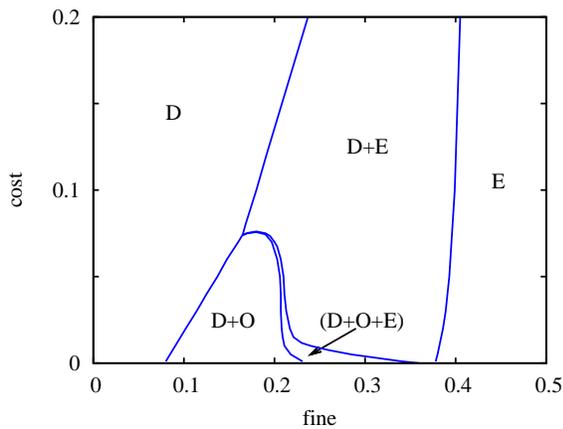,width=7.6cm}}
\caption{(Color online) Phase diagram for the weak peer punishment limit ($m = 1/(G-1)$) at $r=3.5$.}
\label{A_R3_5}
\end{figure}

The general behavior of the four-strategy system at $r=3.5$ is similar to those described above except the missing D+C phase in the low-fine limit. Figure~\ref{A_R3_5} shows that pool punishers can survive with defectors both in the absence or presence of peer punishers at a sufficiently low cost. Otherwise the phase diagram is identical to the result achieved in the absence of pool punisher.

\begin{figure}[ht]
\centerline{\epsfig{file=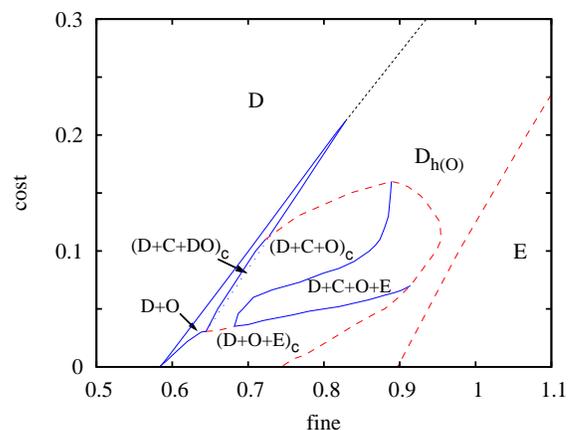,width=7.6cm}}
\caption{(Color online) Cost-fine phase diagram for the weak peer punishment limit ($m = 1/(G-1)$) at $r=2.0$.}
\label{A_R2_0}
\end{figure}

Significantly different and more complex solutions are found at low synergy factor $r$ offering a modest efficiency of investment payed into the common pool. The phase diagram for $r=2$ is plotted in Fig.~\ref{A_R2_0}. In agreement with the previous results some parts of the corresponding phase diagram is identical with those one can obtain if only one type of punishment is allowed. For example, at high fine values, the $E$ strategy conquers not only $D$ but $O$ strategies as well, and the solution reproduces the cases when the player can choose only $D$, $C$, or $E$ strategy. This feature is related to an earlier observation indicating that the increase of fine would not necessarily  help the invasion of $O$ strategy meanwhile peer punishers are unequivocally supported and conquer the system if $\beta$ is enhanced.

On the other hand, one can observe striking similarity with the previous results of a simpler model \cite{szolnoki_pre11} obtained in the absence of peer punishers. This happens in the low-fine region where $E$s cannot fight efficiently against $D$ and die out within a transient period. Accordingly, in this region of the $\beta-\gamma$ plane D+O, (D+C+DO)$_c$, and (D+C+O)$_c$ phases are identified (as detailed in \cite{szolnoki_pre11}) where the subscript "c" refers to self-organizing spatial strategy distribution maintained by cyclic dominance on the analogy of evolutionary rock-paper-scissors games.

The coexistence of both types of punishments occurs in the phases (D+O+E)$_c$ and D+C+O+E indicated in the $\beta-\gamma$ phase diagram (see Fig.~\ref{A_R2_0}). Within the phase (D+O+E)$_c$ three strategies dominate cyclically each other (namely, $D$ beats $E$ beats $O$ beats $D$) and form a self-organizing spatial pattern. At these parameter values the (D+C+O)$_c$ phase is also a possible solutions. 

\subsection{Stability analyses}

\begin{figure}[ht]
\centerline{\epsfig{file=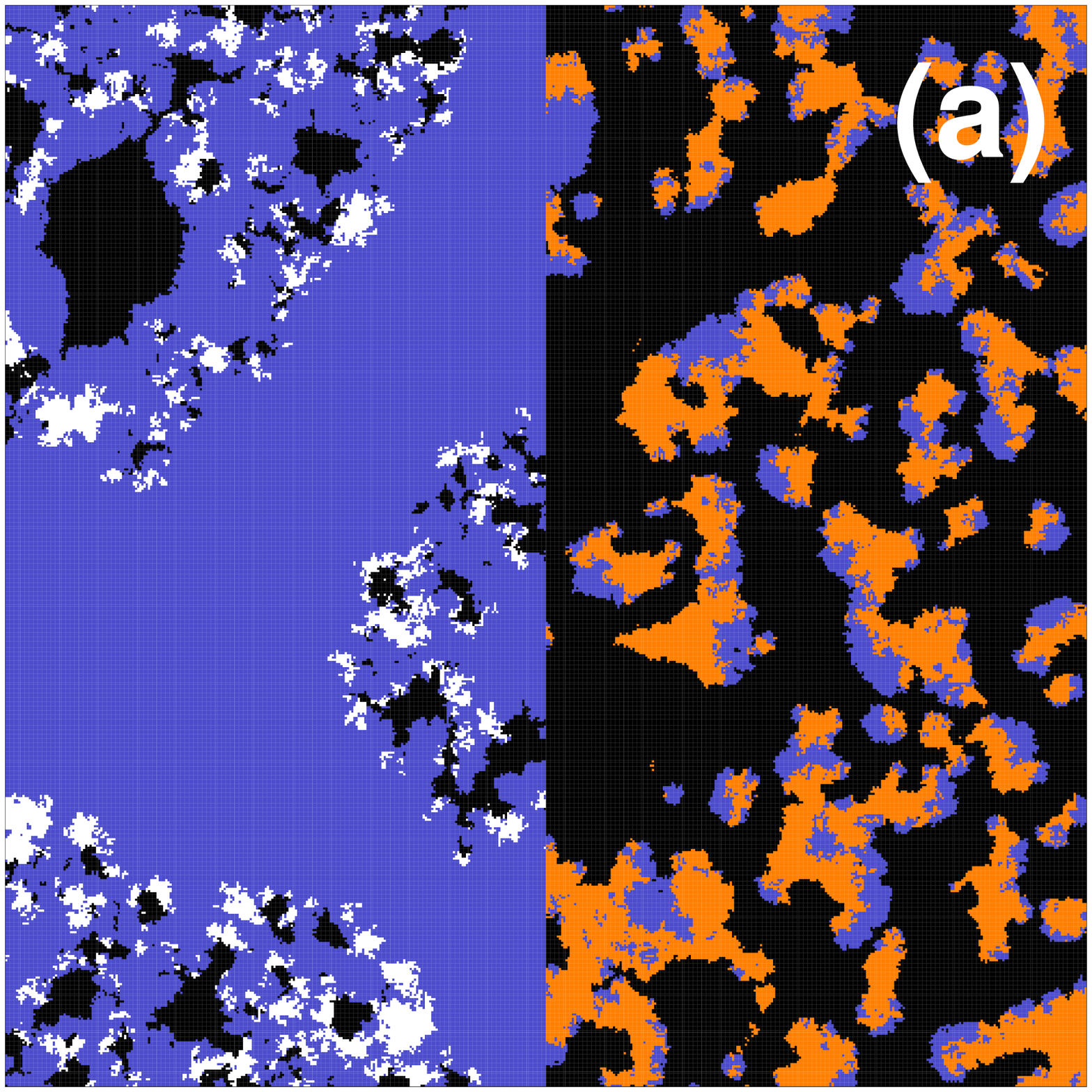,width=4cm}  \epsfig{file=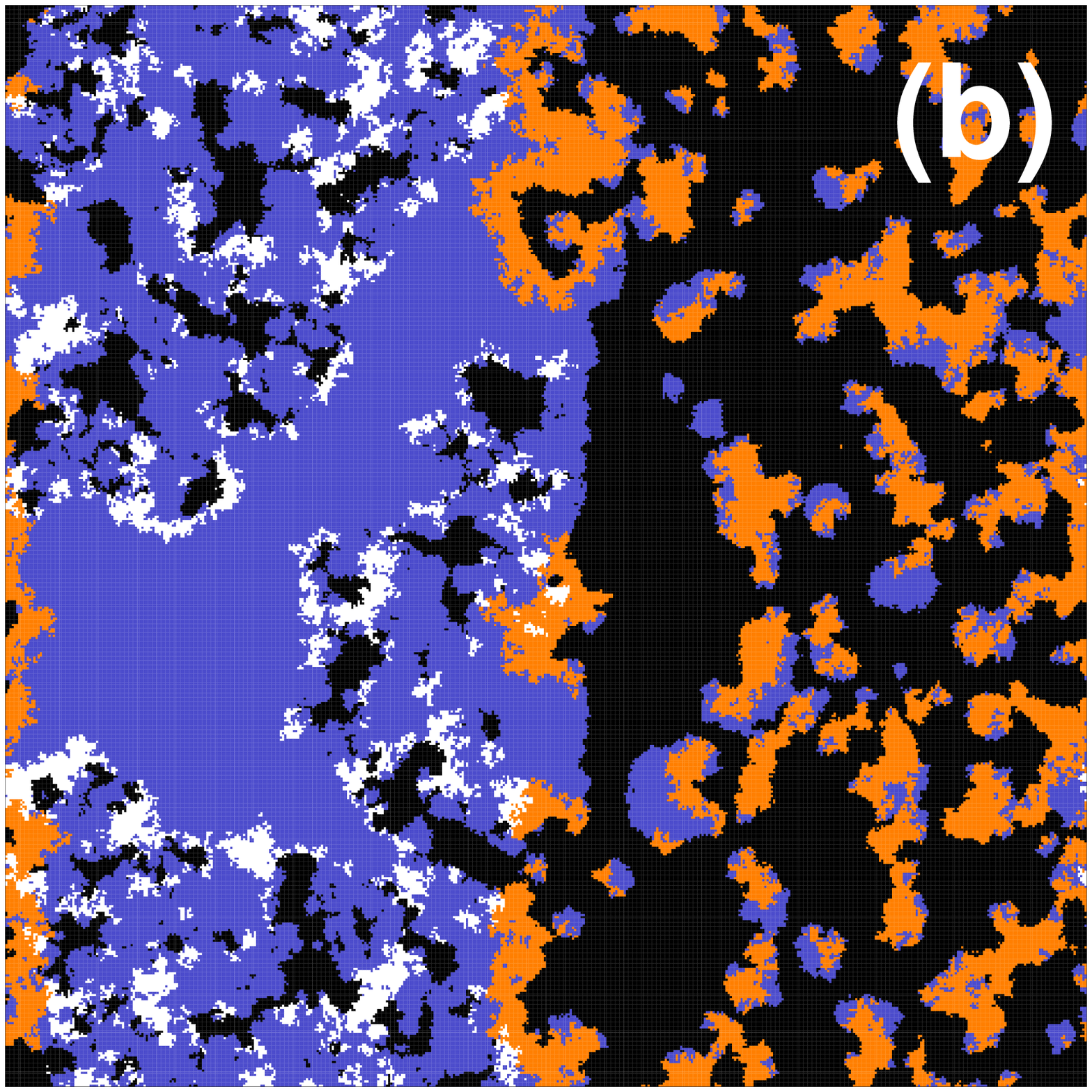,width=4cm}}
\centerline{\epsfig{file=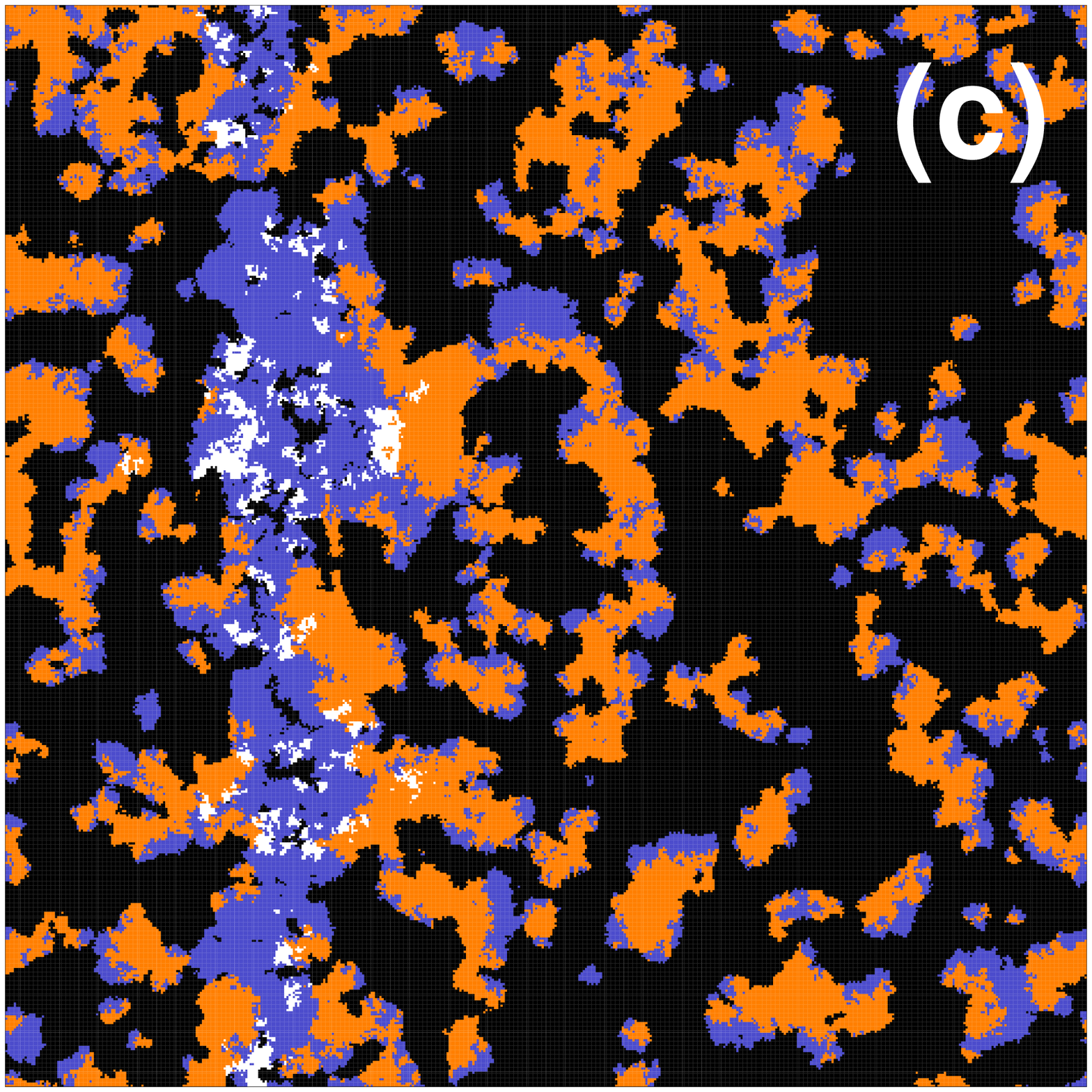,width=4cm}  \epsfig{file=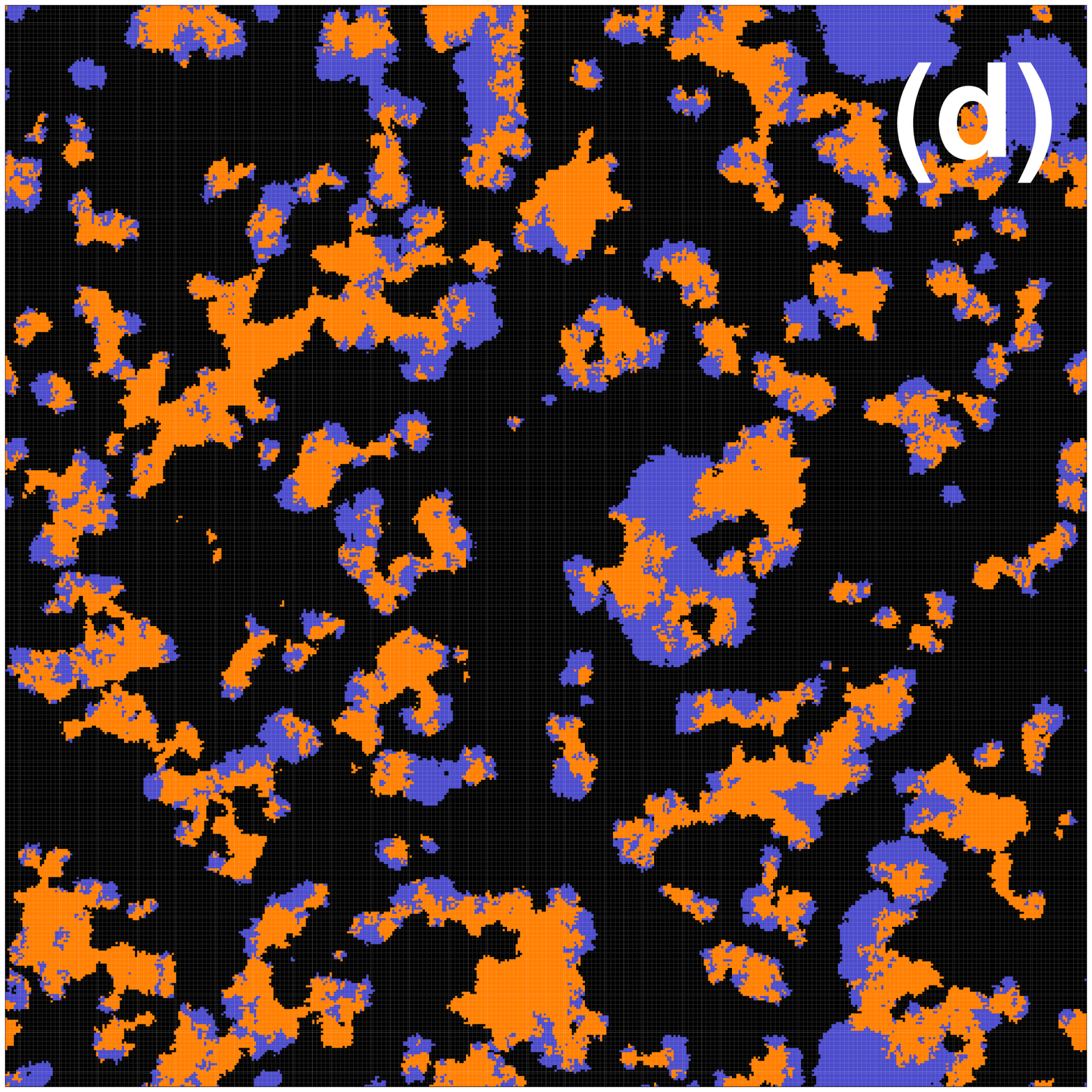,width=4cm}}
\centerline{\epsfig{file=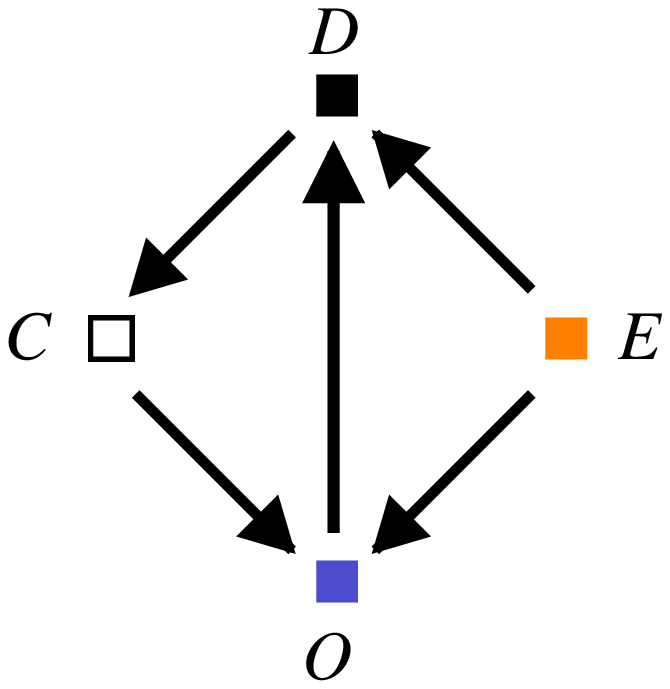,width=4cm}}
\caption{(Color online) Spatial competition between two solutions of three-strategy subsystems 
in the week peer punishment limit ($m = 1/(G-1)$) for
$r=2.0, \beta=0.7$, and $\gamma=0.025$. Before invasions are allowed at $t=0$ MCS along the vertical interfaces, both stationary solutions [(D+C+O)$_c$ (left) and (D+O+E)$_c$ (right)] have been developed without disturbing each other in the corresponding regions. Snapshots of $L=400 \times 400$ part of a $L=800 \times 800$ system are taken at $t=0$ MCS (a), $200$ MCS (b), $1000$ MCS (c), and from the stationary state (d). Lower panel shows the colors of strategies and their relations at these parameters (pointed by an arrow towards the one who is invaded by the other). These are black for $D$, white for $C$, blue (dark gray) for $O$, and orange (light gray) for $E$.}
\label{DOE}
\end{figure}

In the present four-strategy model, however, the (D+O+E)$_c$ coalition (with proper spatio-temporal pattern) is more stable and capable to invade the territory of other solutions as demonstrated by consecutive snapshots in Fig.~\ref{DOE}. For this goal the whole system is divided into large rectangular regions with proper periodic boundary conditions (PBC) for each box during a relaxation time. Within each box only three strategies [$D$+$C$+$O$ or $D$+$O$+$E$] are placed randomly in the initial state. After a suitable relaxation time the proper PBCs are removed and simultaneously the usual PBC is switched on. This trick has allowed us to visualize the spatial competition between the solutions (D+C+O)$_c$ (left) and (D+O+E)$_c$ (right). 

If the system size is large enough then there is always a chance that all the possible solutions can emerge locally somewhere in the system and the most stable solution can finally prevail throughout an invasion process in the whole system. The latter expectation is not necessarily satisfied particularly if the system size is small 
(such as $L<2000$ for $c\approx 0.02$, $r=2$). 
Besides it the "small" size of the system also limits the characteristic size of patterns and prevents the formation of phases including significantly larger correlation lengths. These are the reasons why one cannot achieve reliable MC results on small systems when analyzing the system behavior in the vicinity of a critical point where the correlation length diverges \cite{landau_00}. Further difficulties arise from the fact that the present spatio-temporal patterns can be characterized by two or more length scales  preventing the straightforward application of methods (e.g., finite-size scaling) developed in statistical physics for the investigation of simpler systems  \cite{stanley_71}.

\begin{figure}[ht]
\centerline{\epsfig{file=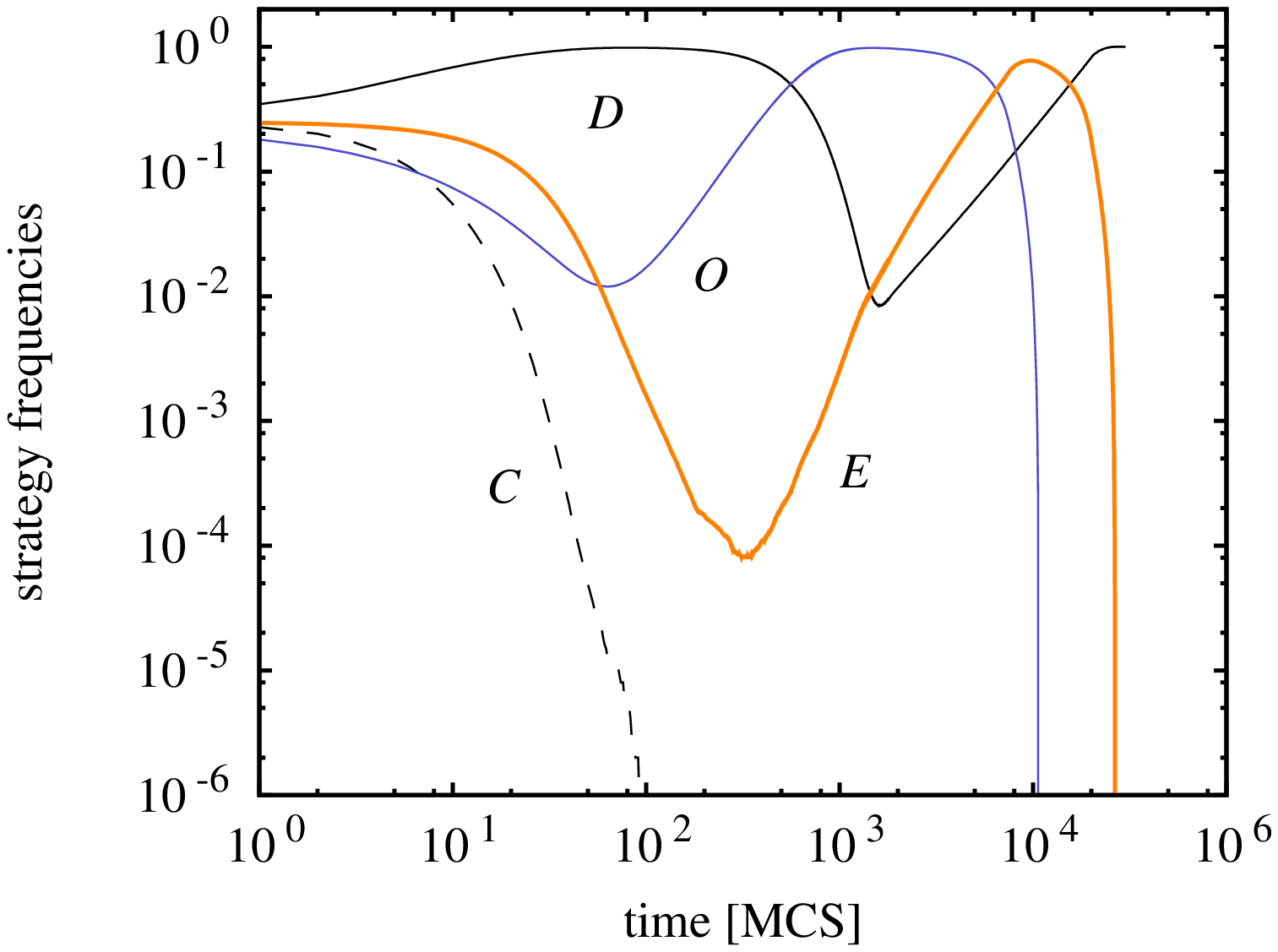,width=7.6cm}}
\centerline{\epsfig{file=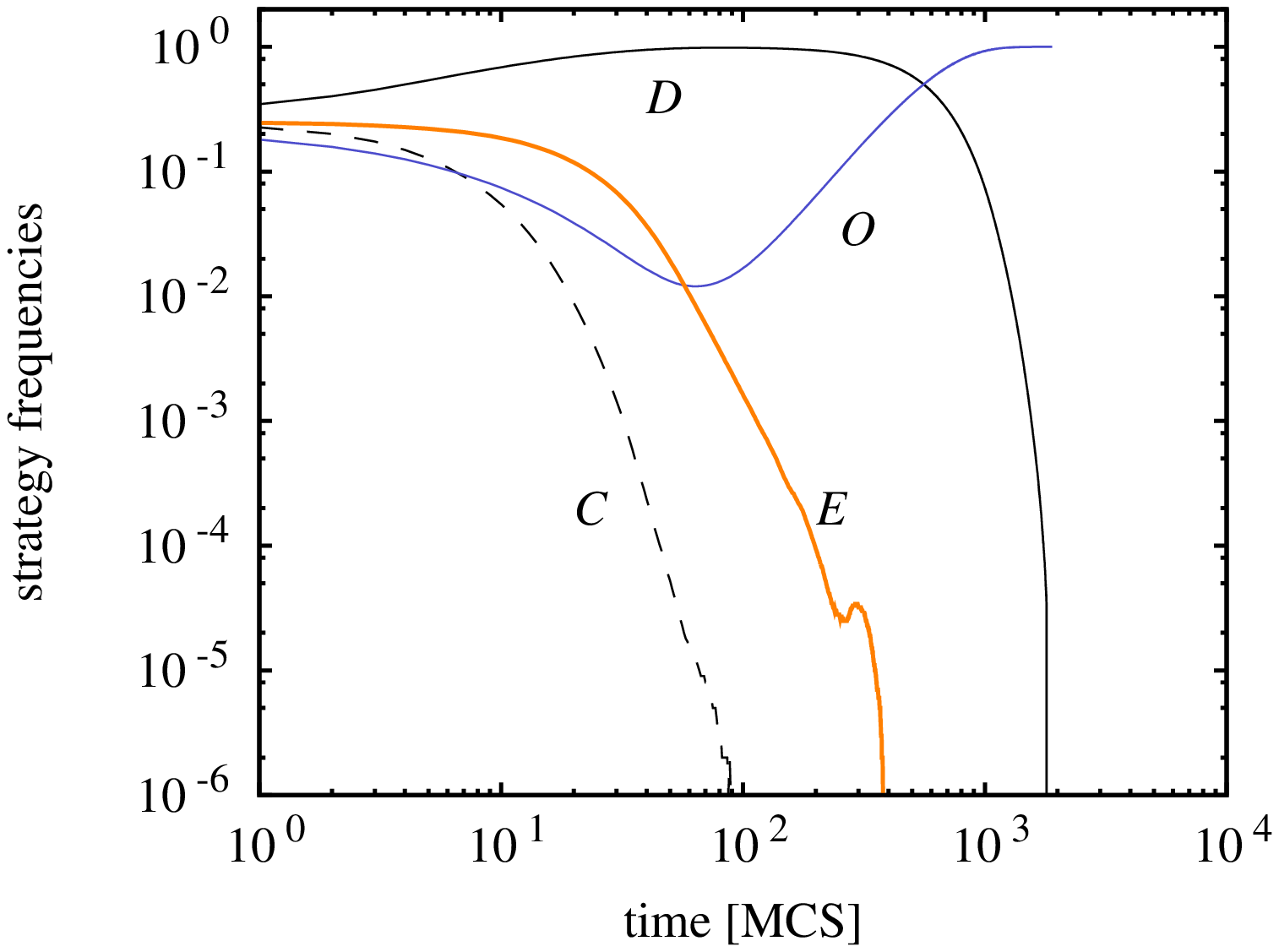,width=7.6cm}}
\centerline{\epsfig{file=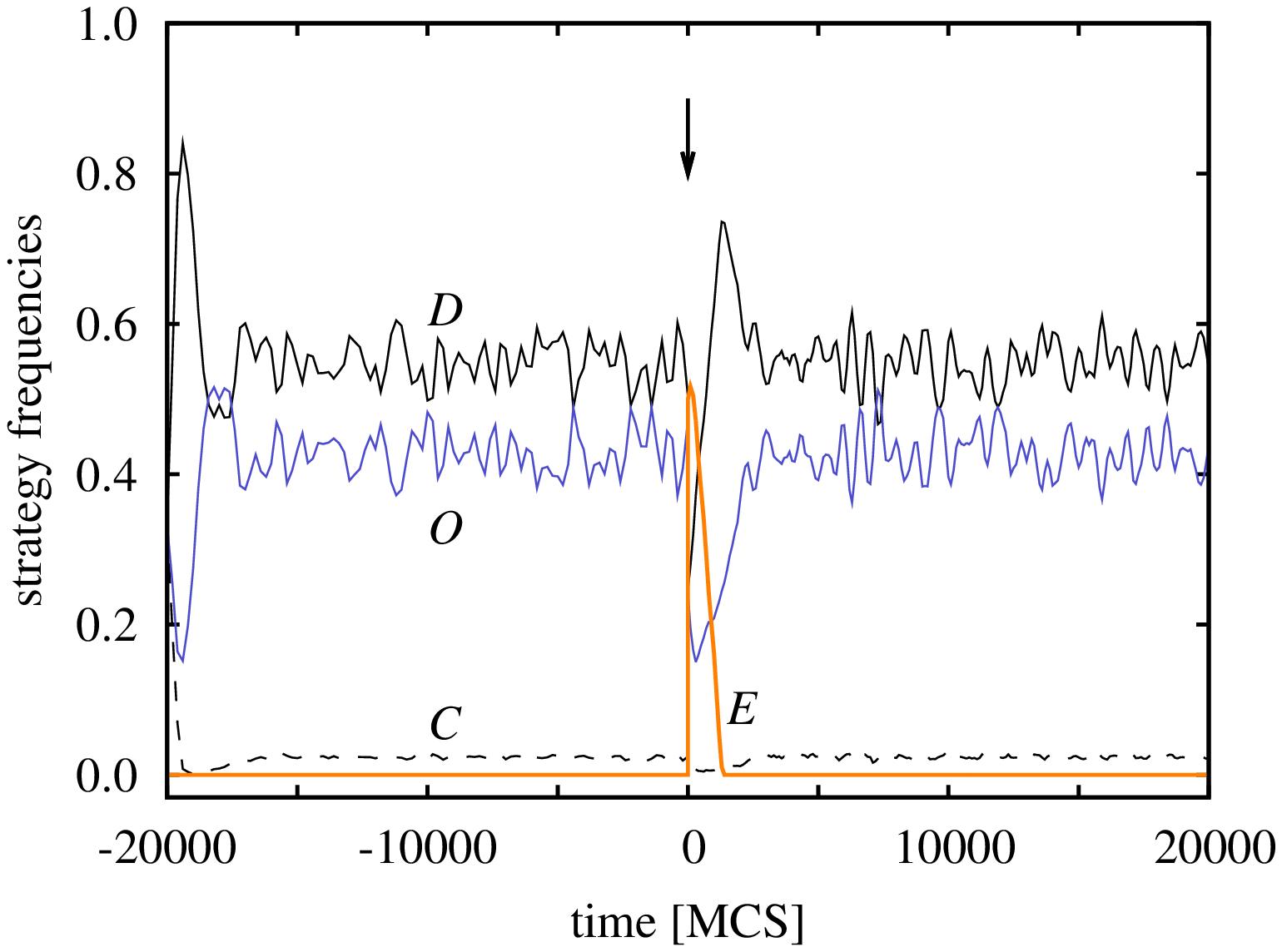,width=7.6cm}}
\caption{(Color online) The upper two plots show evolutionary processes within the region of (D+C+O)$_c$ phase when the system is started from random initial state for $L=5000$,using identical $r=2.0$, $\beta=0.78$, $\gamma=0.1$, and $m=1/(G-1)$ parameter values. The bottom plot demonstrates the stability of (D+C+O)$_c$ phase if we insert a large E domain into the given state at $t=0$ MCS (here $L=1200$).}
\label{random}
\end{figure}

Besides it, the small size decreases the probability of the emergence of phases requiring longer relaxation throughout a complex evolutionary process. Figure~\ref{random} demonstrates the related difficulties of numerical simulations we faced when studying this system for sizes as large as $L=5000$. 
Despite of the large system size the final state is 
still
ambiguous if the system is started from a random initial state. In most cases the system evolves to either D or O state as demonstrated by the upper two plots of Fig.~\ref{random}. Only a very few runs result in a third type (D+C+O)$_c$ phase.  In order to justify the stability of the (D+C+O)$_c$ phase we have performed further stability analyses. Namely, by starting from a three-strategy initial state the stochastic evolution of the (D+C+O)$_c$ phase is interrupted at a time (indicated by an arrow in the bottom plot of Fig.~\ref{random}) and half of the system is replaced by a large domain of E phase and afterwards the simulation is continued. The time-dependence of the strategy frequencies quantify how the original solution is restored. Similar analysis can be done to justify the superiority of (D+C+O)$_c$ phase over the O phase. It is worth mentioning that this ineludible analysis is not time-consuming due to smaller system size used in simulations. 
Furthermore, such a conclusive test cannot be avoided when the model contains more than three competing strategies.

\begin{figure}[ht]
\centerline{\epsfig{file=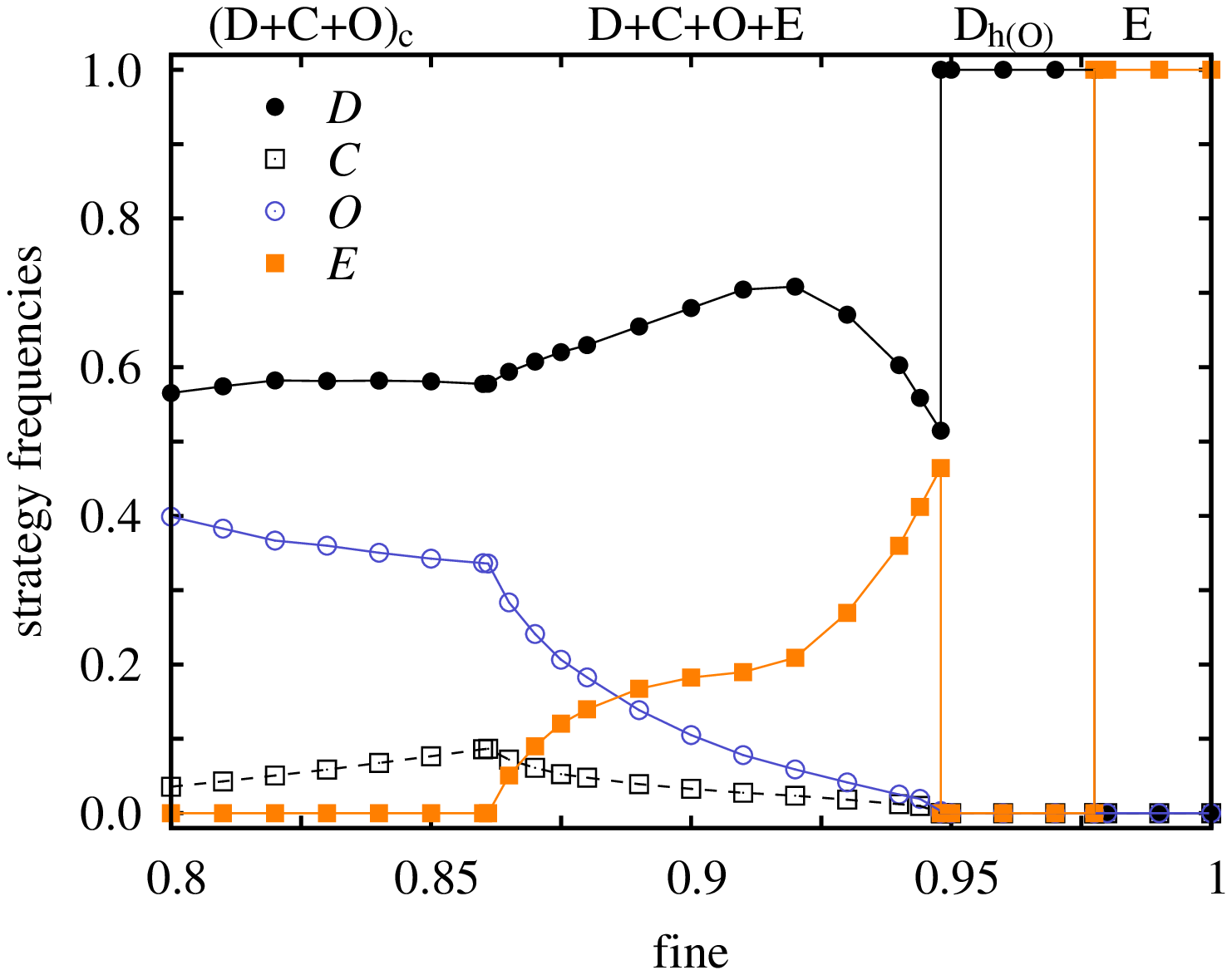,width=7.6cm}}
\centerline{\epsfig{file=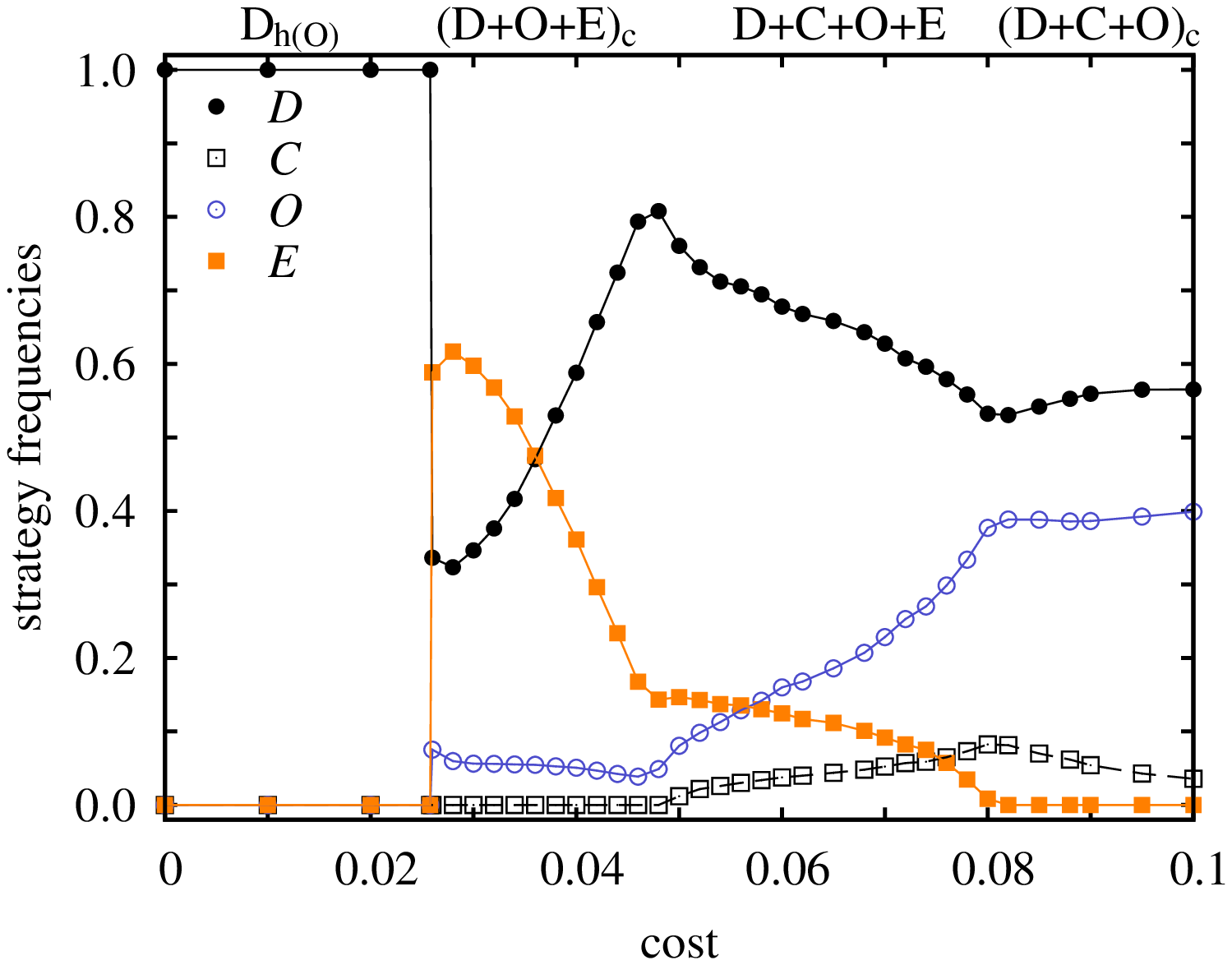,width=7.6cm}}
\caption{(Color online) Strategy frequencies {\it vs}. fine if $\gamma=0.1$ (upper plot) and {\it vs}. cost if $\beta=0.8$ (bottom plot) for $m = 1/(G-1)$ and $r=2.0$. Notation of phases is indicated at the top.}
\label{cross}
\end{figure}

Now we discuss two (perpendicular) cross-sections of the cost-fine phase diagram at $r=2$ (Fig.~\ref{A_R2_0}) where the competition between the two punishing strategies plays relevant role. The upper plot of Fig.~\ref{cross} shows the variation of strategy frequencies in the stationary state when the fine is varied from $\beta=0.8$ to $\beta=1.0$ at a fixed cost. The reader can observe that the four-strategy D+C+O+E phase occurs via a continuous transition from the phase (D+C+O)$_c$ when increasing the fine and subsequently it transforms abruptly into the phase D$_{h(O)}$ where only defectors are present in the final stationary state. As well as previously, the subscript of the notation D$_{h(O)}$ refers to homoclinic instability being different from those discussed in the previous section. 
In the present case the homogeneous D phase can be invaded by the offspring of pool punishers if they help each other by forming a sufficiently large domain. At the same time the growing domain of pool punishers can be eliminated by the offspring of either a single cooperator or peer punisher who is inserted into the territory of pool punishers as a mutant created with an arbitrarily small rate. For both cases defectors play the role of {\it tertius gaudens} and prevail the whole population. For the given cost the peer punishers can beat defectors (with or without the presence of others) if $\beta> 0.978(2)$ when the system evolves into the phase E.

The lower plot of Fig.~\ref{cross} illustrates three consecutive phase transitions when increasing $\gamma$ from 0 to 0.1 for a fixed value of fine ($\beta=0.8$). Notice that within the four strategy phase the frequency of cooperators is low ($\rho_C<0.1$). Despite of the low values of $\rho_C$ the presence of cooperators influences the efficiency of punishing strategies in a complex way indicated by Figs.~\ref{cross}.

\section{Conclusions}
\label{conclusion}

In this work we have compared the efficiency of pool (institutional) and peer (individual) punishments within the framework of spatial public goods game when the strategy evolution is controlled by stochastic imitation (resembling Darwinian selection). 
This study is considered as an initial effort to understand why some societies rely mainly on peer punishment and others prefer pool punishments. As a general conclusion, the output in structured population may depend sensitively on the parameter values characterize the relation of punishment strategies.

Both types of punishment are applied by cooperative players in different ways. The present four-strategy model exhibits a wide variety in the final stationary behavior in the limit of infinitely large system size when tuning the model parameters (synergy factor, cost and fine of punishment) for a fixed level of noise. In many cases the peer punisher strategy seems to be more efficient in the elimination of the "tragedy of the commons" when all players choose defection. The numerical analysis allowed us to identify phases where both types of punishments coexist, sometimes together with the (pure) cooperators weakening the efficiency of punishment. We have found regions in the plane of parameters where the competition between the different punishments helped defectors to prevail the whole system. 

Finally we emphasize some additional and general conclusions extracted during the numerical analysis of the present four-strategy evolutionary game on a square lattice. Namely, we have observed an interesting phase where the spreading of one of the strategies (here cooperation) is favored along an interface and the resultant monolayer can reverse the direction of invasion between the homogeneous domains separated. We think that the structure of the present interactions (namely, the players' income are accumulated from five five-person games) provides convenient conditions for studying these types of self-organizing patterns. Furthermore, we should stress the technical difficulties in the evaluation of phase diagrams describing the boundary between distinguishable stationary behaviors in the limit $L \to \infty$. It turned out that using the concepts of competing associations \cite{szabo_pr07} we should check the direction of invasions between most of the pair of solutions characterizing the spatio-temporal patterns for all possible subsystems if we wish to avoid artifacts related to the complex finite-size effects. At the same time the application of this approach may enhance the accuracy and efficiency of the numerical investigations when quantifying the phase boundaries in the large-size limit. Evidently, the systematic investigation of the finite size effect and also the expansion of a solution in another subsystem solution are inevitable in similar complex systems.

We thank Karl Sigmund for initiating the present investigations and stimulating discussions. This work was supported by the Hungarian National Research Fund (grant K-73449), the Bolyai Research Grant, and the COST Action MP0801 (Physics of Competition and Conflicts).

\end{document}